\documentclass{llncs}
\usepackage{graphics}
\usepackage{amsmath}
\usepackage{graphicx}
\usepackage{amsfonts}
\usepackage{amssymb}

\begin{document}
\title{Nonclassical readout of optical memories under local energy constraint}

\author{Gaetana Spedalieri\inst{1} \and
        Cosmo Lupo\inst{2} \and
        Stefano Pirandola\inst{1}}

\institute{Department of Computer Science, University of York,
York YO10 5GH, UK \and School of Science and Technology,
University of Camerino, Camerino 62032, Italy}

\maketitle

\begin{abstract}
Nonclassical states of light play a central role in many quantum
information protocols. Very recently, their quantum features have
been exploited to improve the readout of information from digital
memories, modeled as arrays of microscopic beam splitters [S.
Pirandola, Phys. Rev. Lett. \textbf{106}, 090504 (2011)]. In this
model of \textquotedblleft quantum reading\textquotedblright, a
nonclassical source of light with Einstein-Podolski-Rosen
correlations has been proven to retrieve more information than any
classical source. In particular, the quantum-classical comparison
has been performed under a global energy constraint, i.e., by
fixing the mean total number of photons irradiated over each
memory cell. In this paper we provide an alternative analysis
which is based on a local energy constraint, meaning that we fix
the mean number of photons \textit{per signal mode} irradiated
over the memory cell. Under this assumption, we investigate the
critical number of signal modes after which a nonclassical source
of light is able to beat any classical source irradiating the same
number of signals.
\end{abstract}

\section{Introduction}

Quantum information has disclosed a modern approach to both
quantum mechanics and information theory \cite{NielsenBook}. Very
recently, this field has been developed into the so-called
\textquotedblleft continuous variable\textquotedblright\ setting,
where information is encoded and processed using quantum systems
with infinite dimensional Hilbert spaces
\cite{RMPWeed,BraREV,BraREV2,GaussSTATES,GaussSTATES2}. Bosonic
systems, such as the radiation modes of the electromagnetic field,
are today the most studied continuous variable systems, thanks to
their strict connection with quantum optics. In the continuous
variable framework, a wide range of results have been successfully
achieved, including quantum
teleportation~\cite{CVtelepo,Bra98,RalphTELE,PirTeleOPMecc,Barlett2003,Sherson}%
, teleportation networks~\cite{TeleNET,teleREV,teleJMO} and
games~\cite{Pirgames,Pir2005}, entanglement swapping
protocols~\cite{Entswap,Entswap2,PirENTswap}, quantum key
distribution~\cite{QKD0,QKD1,Weed,Weed2,Chris}, two-way quantum
cryptography~\cite{PirNATURE,Pir2way}, quantum
computation~\cite{Qcomp1,Qcomp2,Qcomp2b,Qcomp2c,Qcomp2d,Qcomp2e,Qcomp2f} and
cluster quantum computation~\cite{Qcomp3,Qcomp5,Qcomp6,Qcomp7}. Other studies
have lead to the full classification of Gaussian channels and collective
Gaussian attacks~\cite{HolevoCAN,CharacATT,LectNOTES}, the computation of
secret-key capacities and their reverse
counterpart~\cite{Devetak,Deve2,Deve3,PirSKcapacity,RevCOHE}, and possible
schemes for quantum direct
communication~\cite{PirDirectCommunication,PirDcomm2}.

One of the key resources in many protocols of quantum information
is quantum entanglement. In the bosonic setting, quantum
entanglement is usually present under the form of
Einstein-Podolski-Rosen (EPR) correlations~\cite{EPRpaper}, where
the quadrature operators of two separate bosonic modes are so
correlated to beat the standard quantum limit~\cite{note1}. The
simplest source of EPR correlations is the two-mode squeezed
vacuum (TMSV) state. In the number-ket representation this state
is defined by
\[
\left\vert \xi\right\rangle =(\cosh\xi)^{-1}\sum_{n=0}^{\infty}(\tanh\xi
)^{n}\left\vert n\right\rangle _{s}\left\vert n\right\rangle _{i}~,
\]
where $\xi$ is the squeezing parameter and $\{s,i\}$ is an arbitrary pair of
bosonic modes, that we may call \textquotedblleft signal\textquotedblright%
\ and \textquotedblleft idler\textquotedblright. In particular, $\xi$
quantifies the signal-idler entanglement and gives the mean number of photons
\textrm{sinh}$^{2}\xi$ in each mode. Since it is entangled, the TMSV state
cannot be prepared by applying local operations and classical communications
(LOCCs) to a couple of vacua $\left\vert 0\right\rangle _{s}\otimes\left\vert
0\right\rangle _{i}$ or any other kind of tensor product state. For this
reason, the TMSV state cannot be expressed as a classical mixture of coherent
states $\left\vert \alpha\right\rangle _{s}\otimes\left\vert \beta
\right\rangle _{i}$ with $\alpha$ and $\beta$ arbitrary complex amplitudes. In
other words, its P-representation~\cite{Prepres,Prepres2}%
\[
\left\vert \xi\right\rangle \left\langle \xi\right\vert =\int\int d^{2}\alpha
d^{2}\beta\boldsymbol{~}\mathcal{P}(\alpha,\beta)~\left\vert \alpha
\right\rangle _{s}\left\langle \alpha\right\vert \otimes\left\vert
\beta\right\rangle _{i}\left\langle \beta\right\vert ~,
\]
involves a function $\mathcal{P}$ which is non-positive and, therefore, cannot
be considered as a genuine probability distribution. For this reason, the
TMSV\ state is a particular kind of \textquotedblleft
nonclassical\textquotedblright\ state. Other kinds are single-mode squeezed
states and Fock states. By contrast a bosonic state is called
\textquotedblleft classical\textquotedblright\ when its P-representation is
positive, meaning that the state can be written as a classical mixture of
coherent states. Thus a classical source of light is composed by a set of $m$
bosonic modes in a state%
\begin{equation}
\rho=\int d^{2}\alpha_{1}\cdots\int d^{2}\alpha_{m}\boldsymbol{~}%
\mathcal{P}(\alpha_{1},\cdots,\alpha_{m})~\otimes_{k=1}^{m}\left\vert
\alpha_{k}\right\rangle \left\langle \alpha_{k}\right\vert ~, \label{Prepres}%
\end{equation}
where $\mathcal{P}$ is positive and normalized to $1$. Typically, classical
sources are just made by collection of coherent states with amplitudes
$\{\bar{\alpha}_{1},\cdots,\bar{\alpha}_{m}\}$, i.e., $\rho=\otimes_{k=1}%
^{m}\left\vert \bar{\alpha}_{k}\right\rangle \left\langle \bar{\alpha}%
_{k}\right\vert $ which corresponds to have
\[
\mathcal{P}=\prod_{k=1}^{m}\delta^{2}(\alpha_{k}-\bar{\alpha}_{k})~.
\]
In other situations, where the sources are particularly chaotic, they are
better described by a collection of thermal states with mean photon numbers
$\{\bar{n}_{1},\cdots,\bar{n}_{m}\}$, so that%
\[
\mathcal{P}=\prod_{k=1}^{m}\frac{\exp(-\left\vert \alpha_{k}\right\vert
^{2}\bar{n}_{k})}{\pi\bar{n}_{k}}~.
\]
More generally, we can have classical states which are not just tensor
products but they have (classical) correlations among different bosonic modes.

The comparison between classical and nonclassical states has clearly triggered
a lot of interest. The main idea is to compare the use of a candidate
nonclassical state, like the EPR state, with all the classical states for
specific information tasks. One of these tasks has been the detection of
low-reflectivity objects in far target regions under the condition of
extremely low signal-to-noise ratios. This scenario has been called
\textquotedblleft quantum illumination\textquotedblright\ and has been
investigated in a series of papers~\cite{QIll1,QIll2,QIll3,Guha,Devi,YuenNair}.

Most recently, the EPR correlations have been exploited for a completely
different task in a completely different regime of parameters. In the model of
\textquotedblleft quantum reading\textquotedblright\ \cite{QreadingPRL}, the
EPR correlations have been used to retrieve information from digital memories
which are reminiscent of today's optical disks, such as CDs and DVDs. A
digital memory can in fact be modelled as a sequence of cells corresponding to
beam splitters with two possible reflectivities $r_{0}$ and $r_{1}$ (used to
encode a bit of information). By fixing the mean total number of photons
$N$\ irradiated over each memory cell, it is possible to show that a
non-classical source of light with EPR correlations retrieves more information
than any classical source~\cite{QreadingPRL}. In general, the improvement is
found in the regime of few photons ($N=1\div100$)\ and for memories with high
reflectivities, as typical for optical memories. In this regime, the gain of
information given by the quantum reading can be dramatic, i.e., close to $1$
bit for each bit of the memory.

An important point in the study of Ref.~\cite{QreadingPRL} is that the
quantum-classical comparison is performed under a global energy constraint,
i.e., by fixing the total number of photons $N$\ which are irradiated over
each memory cell (on average). Under this assumption, it is possible to
construct an EPR transmitter, made by a suitable number of TMSV states, which
is able to outperform \textit{any} classical source composed by \textit{any}
number of modes. In the following we consider a different and easier
comparison: we fix the number of signal modes irradiated over the target cell
($M$) and the mean number of photons \textit{per signal mode} ($N_{S}$). Under
these assumptions, we compare an EPR transmitter with a classical source.
Then, for fixed $N_{S}$, we determine the critical number of signal modes
$M^{(N_{S})}$ after which an EPR\ transmitter with $M>$ $M^{(N_{S})}$ is able
to beat any classical source (with the same number of signals $M$).

\section{Readout mechanism}

\begin{figure}[ptbh]
\vspace{-1.0cm}
\par
\begin{center}
\includegraphics[width=0.75\textwidth] {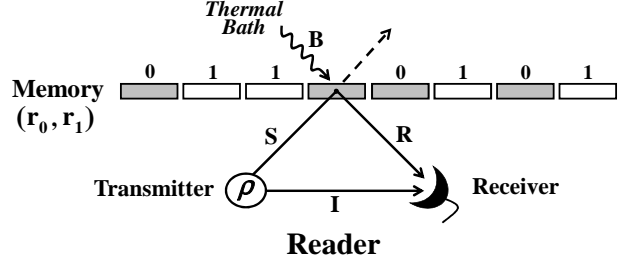}
\end{center}
\par
\vspace{-2.7cm}\caption{\textbf{Model of memory}. Digital
information is stored in a disk whose memory cells are beam
splitter mirrors with different reflectivities: $r=r_{0}$ encoding
bit-value $u=0$ and $r=r_{1}$ encoding bit-value $u=1$.
\textbf{Readout}. A reader is generally composed by a transmitter
and a receiver. It retrieves a stored bit by probing a memory cell
with a signal system $S$ ($M$ bosonic modes) and detecting the
reflected system $R$ together with an idler system $I$ ($L$
bosonic modes). In general, the output system $R$ combines the
signal system $S$ with a bath system $B$ ($M$ bosonic modes in
thermal states). The transmitter is in a state $\rho$ which can be
classical (classical transmitter) or non-classical (quantum
transmitter). In our work, we consider a quantum transmitter with
EPR\ correlations between signal and idler systems.}%
\label{QreadPIC}%
\end{figure}

Here we briefly review the basic readout mechanism of Ref.~\cite{QreadingPRL},
specifying the study to the case of a local energy constraint. Let us consider
a model of digital optical memory (or disk) where the memory cells are beam
splitter mirrors with different reflectivities $r=r_{0},r_{1}$ (with
$r_{1}\geq r_{0}$). In particular, the bit-value $u=0$ is encoded in a
lower-reflectivity mirror ($r=r_{0}$), that we may call a \textit{pit}, while
the bit-value $u=1$ is encoded in a higher-reflectivity mirror ($r=r_{1}$),
that we may call a \textit{land} (see\ Fig.~\ref{QreadPIC}). Close to the
disk, a reader aims to retrieve the value of the bit $u$ which is stored in
each memory cell. For this sake, the reader exploits a transmitter (to probe a
target cell) and a receiver (to measure the corresponding output). In general,
the transmitter consists of two quantum systems, called \textit{signal} $S$
and \textit{idler} $I$, respectively. The signal system $S$ is a set of $M$
bosonic modes which are directly shined on the target cell. The mean total
number of photons of this system is simply given by $N=MN_{S}$, where $N_{S}$
is the mean number of photons per signal mode (simply called \textquotedblleft
energy\textquotedblright, hereinbelow). At the output of the cell, the
reflected system $R$ is combined with the idler system $I$, which is a
supplementary set of bosonic modes whose number $L$ can be completely
arbitrary. Both the systems $R$ and $I$ are finally measured by the receiver
(see\ Fig.~\ref{QreadPIC}).

We assume that Alice's apparatus is very close to the disk, so
that no significant source of noise is present in the gap between the disk and
the decoder. However, we assume that non-negligible noise comes from the
thermal bath present at the other side of the disk. This bath generally
describes stray photons, transmitted by previous cells and bouncing back to
hit the next ones. For this reason, the reflected system $R$ combines the
signal system $S$ with a bath system $B$ of $M$ modes. These environmental
modes are assumed in a tensor product of thermal states, each one with $N_{B}$
mean photons (white thermal noise). In this model we identify five basic
parameters: the reflectivities of the memory $\{r_{0},r_{1}\}$, the
temperature of the bath $N_{B}$, and the profile of the signal $\{M,N_{S}\}$,
which is given by the number of signals $M$\ and the energy $N_{S}$.

In general, for a fixed input state $\rho$ at the transmitter (systems $S,I$),
Alice will get two possible output states $\sigma_{0}$ and $\sigma_{1}$ at the
receiver (systems $R,I$). These output states are the effect of two different
quantum channels, $\mathcal{E}_{0}$ and $\mathcal{E}_{1}$, which depend on the
bit $u=0,1$ stored in the target cell. In particular, we have $\sigma
_{u}=(\mathcal{E}_{u}\otimes\mathcal{I})(\rho)$, where the conditional channel
$\mathcal{E}_{u}$ acts on the signal system, while the identity channel
$\mathcal{I}$ acts on the idler system. More exactly, we have $\mathcal{E}%
_{u}=\mathcal{R}_{u}^{\otimes M}$, where $\mathcal{R}_{u}$ is a one-mode lossy
channel with conditional loss $r_{u}$ and fixed thermal noise $N_{B}$. Now,
the minimum error probability $P_{err}$ affecting the decoding of $u$ is just
the error probability affecting the statistical discrimination of the two
output states, $\sigma_{0}$ and $\sigma_{1}$, via an optimal receiver. This
quantity is equal to $P_{err}=[1-D(\sigma_{0},\sigma_{1})]/2$, where
$D(\sigma_{0},\sigma_{1})$ is the trace distance between $\sigma_{0}$ and
$\sigma_{1}$~\cite{Helstrom,Fuchs,FuchsThesis}. Clearly, the value of
$P_{err}$ determines the average amount of information which is decoded for
each bit stored in the memory. This quantity is equal to $J=1-H(P_{err})$,
where $H(x):=-x\log_{2}x-(1-x)\log_{2}(1-x)$ is the usual formula for the
binary Shannon entropy. In the following, we compare the performance of
decoding in two paradigmatic situations, one where the transmitter is
described by a non-classical state (quantum transmitter) and one where the
transmitter is in a classical state (classical transmitter). In particular, we
show how a quantum transmitter with EPR correlations (EPR transmitter) is able
to outperform classical transmitters. The quantum-classical comparison is
performed for fixed signal profile $\{M,N_{S}\}$. Then, for various fixed
values of the energy $N_{S}$ (local energy constraint), we study the critical
number of signal modes  $M^{(N_{S})}$ after which an EPR transmitter (with
$M>M^{(N_{S})}$ signals) is able to beat any classical transmitter (with the
same number of signals $M$).

\section{Quantum-classical comparison}

First let us consider a classical transmitter. A classical transmitter with
$M$ signals and $L$ idlers is described by a classical state $\rho$ as
specified by Eq.~(\ref{Prepres}) with $m=M+L$. In other words it is a
probabilistic mixture of multi-mode coherent states $\otimes_{k=1}%
^{M+L}\left\vert \alpha_{k}\right\rangle \left\langle \alpha_{k}\right\vert $.
Given this transmitter, we consider the corresponding error probability
$P_{err}^{class}$ which affects the readout of the memory. Remarkably, this
error probability is lower-bounded by a quantity which depends on the signal
profile $\{M,N_{S}\}$, but not from the number $L$ of the idlers and the
explicit expression of the $\mathcal{P}$-function. In fact, we
have~\cite{QreadingPRL}%
\begin{equation}
P_{err}^{class}\geq\mathcal{C}(M,N_{S}):=\frac{1-\sqrt{1-F(N_{S})^{M}}}{2}~,
\label{CB_cread}%
\end{equation}
where $F(N_{S})$ is the fidelity between $\mathcal{R}_{0}(|N_{S}^{1/2}%
\rangle\langle N_{S}^{1/2}|)$ and $\mathcal{R}_{1}(|N_{S}^{1/2}\rangle\langle
N_{S}^{1/2}|)$, the two possible outputs of the single-mode coherent state
$|N_{S}^{1/2}\rangle\langle N_{S}^{1/2}|$. As a consequence, all the classical
transmitters with signal profile $\{M,N_{S}\}$ retrieve an information which
is upper-bounded by $J_{class}:=1-H[\mathcal{C}(M,N_{S})]$.

Now, let us construct a transmitter having the same signal profile
$\{M,N_{S}\}$, but possessing EPR correlations between signals and idlers.
This is realized by taking $M$ identical copies of a TMSV state, i.e.,
$\rho=\left\vert \xi\right\rangle \left\langle \xi\right\vert ^{\otimes M}$
where $N_{S}=\mathrm{sinh}^{2}\xi$. Given this transmitter, we consider the
corresponding error probability $P_{err}^{quant}$ affecting the readout of the
memory. This quantity is upper-bounded by the quantum Chernoff bound
\cite{QCbound,QCbound2,QCbound3,QCbound4,MinkoPRA}%
\begin{equation}
P_{err}^{quant}\leq\mathcal{Q}(M,N_{S}):=\frac{1}{2}\left[  \inf_{s\in
(0,1)}\mathrm{Tr}(\theta_{0}^{s}\theta_{1}^{1-s})\right]  ^{M},
\label{QCB_qread}%
\end{equation}
where $\theta_{u}:=(\mathcal{R}_{u}\otimes\mathcal{I})(\left\vert
\xi\right\rangle \left\langle \xi\right\vert )$. Since $\theta_{0}$ and
$\theta_{1}$ are Gaussian states, we can write their symplectic
decompositions~\cite{Alex} and compute the quantum Chernoff bound using the
general formula for multimode Gaussian states given in Ref.~\cite{MinkoPRA}.
Then, we can easily compute a lower bound $J_{quant}:=1-H[\mathcal{Q}%
(M,N_{S})]$ for the information which is decoded via this quantum transmitter.

In order to show an improvement with respect to the classical case, it is
sufficient to prove the positivity of the \textquotedblleft information
gain\textquotedblright\ $G:=J_{quant}-J_{class}$. This quantity is in fact a
lower bound for the average information which is gained by using the EPR
quantum transmitter instead of every classical transmitter. Roughly speaking,
the value of $G$ estimates the minimum information which is gained by the
quantum readout for each bit of the memory. In general, $G$ is a function of
all the basic parameters of the model, i.e., $G=G(M,N_{S},r_{0},r_{1},N_{B})$.
Numerically, we can easily find signal profiles $\{M,N_{S}\}$, classical
memories $\{r_{0},r_{1}\}$, and thermal baths $N_{B}$, for which we have the
quantum effect $G>0$. Some of these values are reported in the following
table.%
\[%
\begin{tabular}
[c]{|c|c|c|c|c|c|}\hline
$~M~$ & $~~N_{S}~~$ & $~~~~r_{0}~~~~$ & $~~~~r_{1}~~~~$ & $~~N_{B}~~$ &
$~~~G~($bits$)~~~$\\\hline
$1$ & $3.5$ & $0.5$ & $0.95$ & $0.01$ & $~6.2\times10^{-3}$\\\hline
$10$ & $1$ & $0.2$ & $0.8$ & $0.01$ & $~3.4\times10^{-2}$\\\hline
$30$ & $1$ & $0.38$ & $0.85$ & $1$ & $~1.2\times10^{-3}$\\\hline
$100$ & $0.1$ & $0.25$ & $0.85$ & $0.01$ & $~5.9\times10^{-2}$\\\hline
$200$ & $0.1$ & $0.6$ & $0.95$ & $0.01$ & $0.22$\\\hline
$2\times10^{5}$ & $0.01$ & $0.995$ & $1$ & $0$ & $0.99$\\\hline
\end{tabular}
\ \ \ \
\]
Note that we can find choices of parameters where $G\simeq1$, i.e., the
classical readout of the memory does not decode any information whereas the
quantum readout is able to retrieve all of it. As shown in the last row of the
table, this situation can occur when the both the reflectivities of the memory
are very close to $1$. From the first row of the table, we can acknowledge
another remarkable fact: for a land-reflectivity $r_{1}$\ sufficiently close
to $1$, one signal with few photons can give a positive gain. In other words,
the use of a single, but sufficiently entangled, TMSV state $\left\vert
\xi\right\rangle \left\langle \xi\right\vert $ can outperform every classical
transmitter, which uses one signal mode with the same energy (and potentially
infinite idler modes).

\begin{figure}[ptbh]
\vspace{-0.4cm}
\par
\begin{center}
\includegraphics[width=0.55\textwidth] {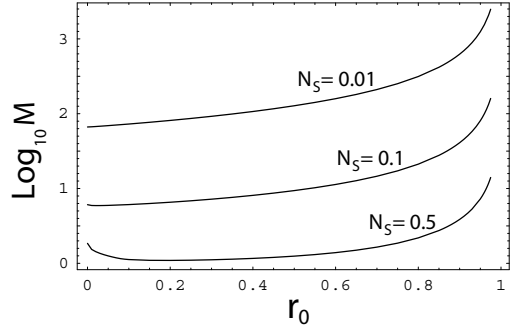}
\end{center}
\par
\vspace{-0.6cm}\caption{Number of signals $M$ (logarithmic scale)\
versus pit-reflectivity $r_{0}$. The curves refer to $N_{S}=0.01$,
$0.1$ and $0.5$ photons. For each value of the energy $N_{S}$, we
plot the critical number $M^{(N_{S})}(r_{0})$ as function of
$r_{0}$. All the curves have an asymptote at $r_{0}=1$. For
$N_{S}\gtrsim2.5$ photons (curves not shown), we have
another asymptote at $r_{0}=0$. }%
\label{PRLmin}%
\end{figure}

According to our numerical investigation, the quantum readout is
generally more powerful when the land-reflectivity is sufficiently
high (i.e., $r_{1}\gtrsim0.8$). For this reason, it is very
important to analyze the scenario in the limit of ideal
land-reflectivity ($r_{1}=1$). Let us call \textquotedblleft ideal
memory\textquotedblright\ a classical memory with $r_{1}=1$.
Clearly, this memory is completely characterized by the value of
its pit-reflectivity $r_{0}$. For ideal memories, the quantum
Chernoff bound of Eq.~(\ref{QCB_qread}) takes the analytical form
\[
\mathcal{Q}=\frac{1}{2}\{[1+(1-\sqrt{r_{0}})N_{S}]^{2}+N_{B}(2N_{S}%
+1)(1-r_{0})\}^{-M},
\]
and the classical bound of Eq.~(\ref{CB_cread}) can be computed
using
\[
F(N_{S})=\gamma^{-1}\exp[-\gamma^{-1}(1-\sqrt{r_{0}})^{2}N_{S}]~,
\]
where $\gamma:=1+(1-r_{0})N_{B}$~\cite{QreadingPRL}. Using these
formulas, we can study the behavior of the gain $G$ in terms of
the remaining parameters $\{M,N_{S},r_{0},N_{B}\}$. Let us
consider an ideal memory with generic $r_{0}\in\lbrack0,1)$ in a
generic thermal bath $N_{B}\geq0$. For a fixed energy $N_{S}$, we
consider the minimum number of signals $M^{(N_{S})}$ above which
$G>0$~\cite{note2}. This critical number can be defined
independently from the thermal noise $N_{B}$ (via an implicit
maximization over $N_{B}$). Then, for a given value of the energy
$N_{S}$, the critical
number $M^{(N_{S})}$ is a function of $r_{0}$ alone, i.e., $M^{(N_{S}%
)}=M^{(N_{S})}(r_{0})$. Its behavior is shown in Fig.~\ref{PRLmin}
for different values of the energy. 

It is remarkable that, for low-energy signals ($N_{S}=0.01\div1$ photons), the
critical number $M^{(N_{S})}(r_{0})$ is finite for every $r_{0}\in\lbrack
0,1)$. This means that, for ideal memories and low-energy signals, there
always exists a finite number of signals $M^{(N_{S})}$ above which the quantum
readout of the memory is more efficient than its classical readout. In other
words, there is an EPR\ transmitter with $M>M^{(N_{S})}$ able to beat any
classical transmitter with the same number of signals $M$. In the considered
low-energy regime, $M^{(N_{S})}(r_{0})$ is relatively small for almost all the
values of $r_{0}$, except for $r_{0}\rightarrow1$\ where $M^{(N_{S})}%
(r_{0})\rightarrow\infty$. In fact, for $r_{0}\simeq1$, we derive $M^{(N_{S}%
)}(r_{0})\simeq\lbrack4N_{S}(2N_{S}+1)(1-r_{0})]^{-1}$, which diverges at
$r_{0}=1$. Such a divergence is expected, since we must have $P_{err}%
^{quant}=P_{err}^{class}=1/2$\ for $r_{0}=r_{1}$ (see
Appendix~\ref{app1} for details). Apart from the divergence at
$r_{0}=1$, in all the other points $r_{0}\in\lbrack0,1)$, the
critical number $M^{(N_{S})}(r_{0})$ decreases for increasing
energy $N_{S}$ (see Fig.~\ref{PRLmin}). In particular, for
$N_{S}=1$ photon, we have $M^{(N_{S})}(r_{0})\simeq1$\ for most of
the reflectivities $r_{0}$. In other words, for energies around
one photon, a single TMSV state is sufficient to provide a
positive gain for most of the ideal memories. However, the
decreasing trend of $M^{(N_{S})}(r_{0})$ does not continue for
higher energies ($N_{S}\geq1$). In fact, just after $N_{S}=1$,
$M^{(N_{S})}(r_{0})$ starts to increase around $r_{0}=0$. In
particular, for
$N_{S}\geq1$, we can derive $M^{(N_{S})}(0)\simeq(\ln2)[2\ln(1+N_{S}%
)-N_{S}]^{-1}$, which is increasing in $N_{S}$, and becomes infinite at
$N_{S}\simeq2.5$. As a consequence, for $N_{S}\gtrsim2.5$ photons, we have a
second asymptote appearing at $r_{0}=0$ (see Appendix~\ref{app2} for details).
This means that the use of high-energy signals ($N_{S}\gtrsim2.5$) does not
assure positive gains for memories with extremal reflectivities $r_{0}=0$ and
$r_{1}=1$.

\section{Conclusion}

In conclusion, we have considered the basic model of digital memory studied in
Ref.~\cite{QreadingPRL}, which is composed of beam splitter mirrors with
different reflectivities. Adopting this model, we have compared an EPR
transmitter with classical sources for fixed signal profiles, finding positive
information gains for memories with high land-reflectivities ($r_{1}%
\gtrsim0.8$). Analytical results can be derived in the limit of ideal
land-reflectivity ($r_{1}=1$) which defines the regime of ideal memories. In
this case, by fixing the mean number of photons per signal mode (local energy
constraint), we have computed the critical number of signals after which an
EPR\ transmitter gives positive information gains. For low-energy signals
($0.01\div1$ photons) this critical number is finite and relatively small for
every ideal memory. In particular, an EPR\ transmitter with one TMSV state can
be sufficient to achieve positive information gains for almost all the ideal
memories. Finally, our results corroborate the outcomes of
Ref.~\cite{QreadingPRL} providing an alternative study which considers a local
energy constraint instead of a global one. As discussed in
Ref.~\cite{QreadingPRL}, potential applications are in the technology of
optical digital memories and involve increasing their data-transfer rates and
storage capacities.

\appendix

\section{General asymptote at $r_{0}=1$\label{app1}}

According to Fig.~\ref{PRLmin}, the critical number $M^{(N_{S})}(r_{0})$
diverges for $r_{0}\rightarrow1$. Let us analyze the behavior of $G$ around
the singular point $r_{0}=1$, by setting $r_{0}=1-\varepsilon$ and expanding
$G$ for $\varepsilon\rightarrow0^{+}$. It is easy to check that, for every
$N_{B}$, we have $G>0$ if and only if $M>[4N_{S}(2N_{S}+1)\varepsilon]^{-1}$.
In particular, in the absence of thermal noise ($N_{B}=0$), we have%
\begin{equation}
G=\frac{MN_{S}(4MN_{S}-1)\varepsilon^{2}}{8\ln2}+O(\varepsilon^{3})~,
\label{G_expans}%
\end{equation}
which is positive if and only if $M>(4N_{S})^{-1}$.

\textbf{Proof.}~~Note that $G>0$ if and only if $\Delta:=\mathcal{Q}%
(M,N_{S})-\mathcal{C}(M,N_{S})<0$. Let us expand $\Delta=\Delta(M,N_{S}%
,N_{B},1-\varepsilon)$\ at the first order in $\varepsilon$. For a given
$N_{B}>0$, we have $\Delta=\frac{1}{2}\left[  \left(  MN_{B}\varepsilon
\right)  ^{1/2}-M(N_{B}+N_{S}+2N_{B}N_{S})\varepsilon\right]  +O(\varepsilon
^{3/2})$, which is negative if and only if%
\[
M>\frac{N_{B}}{(N_{B}+N_{S}+2N_{B}N_{S})^{2}\varepsilon}:=\kappa(N_{B})~.
\]
Notice that $\kappa(N_{B})$ is maximum for $N_{B}^{\ast}=N_{S}(1+2N_{S})^{-1}%
$. Then, for every $N_{B}>0$, we have $\Delta<0$ if and only if%
\begin{equation}
M>\kappa(N_{B}^{\ast})=\frac{1}{4N_{S}(2N_{S}+1)\varepsilon}~. \label{Bound_M}%
\end{equation}
Now, let us consider the particular case of $N_{B}=0$. In this case, we have
the first-order expansion $\Delta=\left(  MN_{S}\right)  ^{1/2}[1-2\left(
MN_{S}\right)  ^{1/2}]\varepsilon/4+O(\varepsilon^{2})$, or equivalently the
second-order expansion of $G$ given in Eq.~(\ref{G_expans}). It is clear that
$\Delta<0$, i.e., $G>0$, when $M>1/4N_{S}$. However, this condition is less
restrictive than the one in Eq.~(\ref{Bound_M}) which, therefore, can be
extended to every $N_{B}\geq0$.~$\blacksquare$

\section{High-energy asymptote at $r_{0}=0$\label{app2}}

\begin{figure}[ptbh]
\vspace{-0.4cm}
\par
\begin{center}
\includegraphics[width=0.37\textwidth] {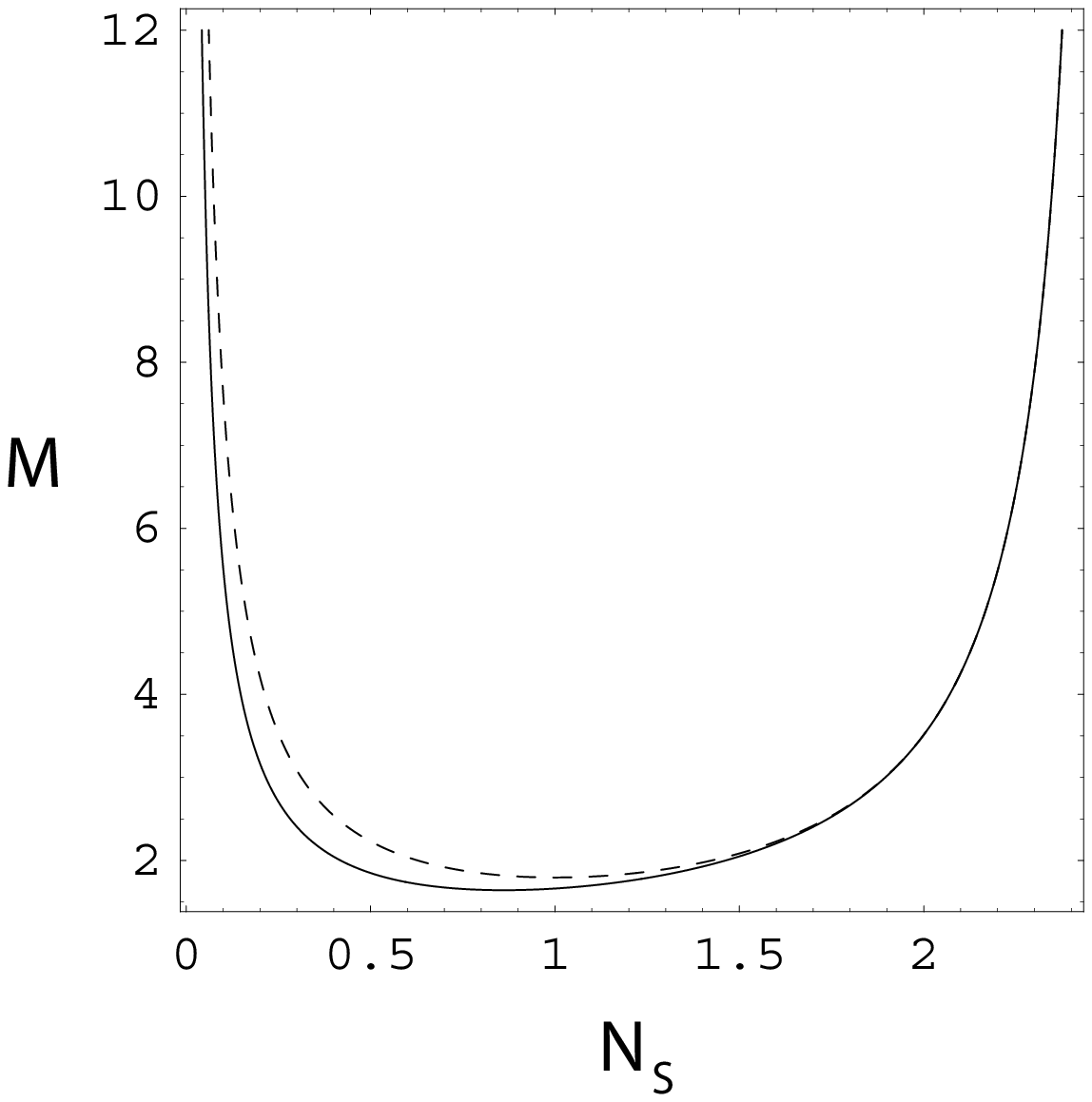}
\end{center}
\par
\vspace{-0.7cm}\caption{Minimum number of signals $M$ versus
energy $N_{S}$. The solid curve represents $M^{(N_{S})}(0)$ while
the dashed curve represents $\tilde{M}$. Note that the minimum
number of signals is actually given by $\left\lceil M\right\rceil
$ where $\left\lceil \cdots\right\rceil $ is the
ceiling function.}%
\label{ApproxPIC}%
\end{figure}

Let us analyze the behavior of $M^{(N_{S})}(r_{0})$ for $N_{S}\geq1$ and
$r_{0}=0$. One can check that, for $N_{S}\geq1$, the greatest value of
$M^{(N_{S})}(0)$\ occurs when $N_{B}=0$. In this case, i.e., for $r_{0}%
=N_{B}=0$ and $r_{1}=1$, we have $\mathcal{Q}(M,N_{S})=[\left(  1+N_{S}%
\right)  ^{-2M}]/2$, and%
\[
\mathcal{C}(M,N_{S})=\frac{1-\sqrt{1-e^{-MN_{S}}}}{2}\overset{M\gg
1}{\longrightarrow}\frac{e^{-MN_{S}}}{4}:=\mathcal{C}^{\infty}.
\]
Let us consider the critical value $M^{(N_{S})}(0)$ of $M$ such that
$G(M,N_{S})=0$, which is equivalent to $\mathcal{Q}=\mathcal{C}$. We also
consider the value $\tilde{M}$ such that $\mathcal{Q}=\mathcal{C}^{\infty}$.
Since $\mathcal{C}^{\infty}\leq\mathcal{C}$, we have that $M^{(N_{S})}%
(0)\geq\tilde{M}$. Actually, we find that $M^{(N_{S})}(0)\simeq\tilde{M}$ with
very good approximation when $N_{S}\geq1$ (see Fig.~\ref{ApproxPIC}). Then,
for every $N_{S}\geq1$, we can set $M^{(N_{S})}(0)\simeq(\ln2)\left[
2\ln(1+N_{S})-N_{S}\right]  ^{-1}$. The latter quantity becomes infinite for
$2\ln(1+N_{S})=N_{S}$, i.e., for $N_{S}\gtrsim2.51$ photons.

\begin{thebibliography}{99}
\bibitem{note1}For two bosonic modes, $A$ and $B$, with quadratures $\hat{q}_{A}$,
$\hat{p}_{A}$, $\hat{q}_{B}$ and $\hat {p}_{B}$, one can define
the two operators $\hat{q}_{-}:=(\hat{q}_{A}-\hat
{q}_{B})/\sqrt{2}$ (relative position)\ and
$\hat{p}_{+}:=(\hat{p}_{A}+\hat {p}_{B})/\sqrt{2}$ (total
momentum). Then, the system has EPR correlations (in these
operators) if $V(\hat{q}_{-})+V(\hat{p}_{+})<2\nu_{0}$, where
$V(\cdot)$ is the variance, and $\nu_{0}$ is the standard quantum
limit ($\nu_{0}=1$ in this paper.) A bosonic system with EPR
correlations is entangled, but the contrary is not necessarily
true. In today's quantum optics labs, the EPR\ correlations
represent the standard type of continuous variable entanglement.
These correlations are usually generated via spontaneous
parametric down conversion.
\bibitem{note2}To be precise, the critical number $M^{(N_{S})}$
is a solution of the equation $G=0$. From this real value we
derive the minimum number of signals (which is an integer) by
taking its ceiling function $\lceil M^{(N_{S})}\rceil$.
\end{thebibliography}
\end{document}